\input psfig.tex
\documentstyle[12pt,world_sci]{article}
\def\NPB#1{{\em Nucl.~Phys.~}{\bf B#1}}
\def\PLB#1{{\em Phys.~Lett.~}{\bf B#1}}
\def\ARNPS#1{{\em Ann.~Rev.~Nucl.~Part.~Sci.~}{\bf #1}}

\def\Zbb{Zb\bar{b}}
\def\glb{g_L^b}
\def\grb{g_R^b}
\def\dglb{\delta g_L^b}
\def\dgrb{\delta g_R^b}
\def\ds{\delta s^2}

\def\drho{\delta\rho}
\def\seff{\sin^2\theta_{\rm eff}}
\def\sig0{\sigma_{\rm had}^0}

\def\Gamb{\Gamma_{b\bar{b}}}

\def\xib{\xi_b}
\def\zetab{\zeta_b}

\def\beq{\begin{equation}}
\def\eeq{\end{equation}}
\def\beqa{\begin{eqnarray}}
\def\eeqa{\end{eqnarray}}

\begin{document}

\begin{flushleft}
Presented at the DPF'94 Meeting  \hfill  FERMILAB-CONF-94/279-T \\
Albuquerque, NM, Aug. 2--6, 1994 \hfill  EFI 94-44 \\
Proceedings to be published by World Scientific \hfill August 1994 \\
\end{flushleft}

\title{{\bf AN ANALYSIS OF NON--OBLIQUE CORRECTIONS TO\\
           THE $\Zbb$ VERTEX \\}}
\author{TATSU TAKEUCHI\thanks{Presenting author} \\
{\em Fermi National Accelerator Laboratory \\
     P.O. Box 500, Batavia, IL 60510}\\
\vspace{0.3cm}
and\\
\vspace*{0.3cm}
AARON K. GRANT and JONATHAN L. ROSNER\\
{\em Enrico Fermi Institute and Department of Physics,
University of Chicago \\
5640 S. Ellis Avenue, Chicago, IL 60637}}

\maketitle

\setlength{\baselineskip}{2.6ex}

\begin{center}
\parbox{13.0cm}
{\begin{center} ABSTRACT \end{center}
{\small \hspace*{0.3cm}
We present a model--independent analysis of the $Z b \bar b$ vertex,
with the aim of constraining contributions of new physics to the
left- and right--handed couplings of the $b$.  We find that the left--handed
coupling of the $b$ is quite narrowly constrained by present data, but that
the right--handed coupling is still largely unconstrained.}}
\end{center}

\section{Introduction}

Recently there has been increasing interest in extensions of the
Standard Model (SM) which predict sizable corrections to the $\Zbb$ vertex.
This interest is motivated in part
by the fact that a deviation from the SM prediction of
$R_b = \Gamma_{b\bar{b}}/\Gamma_{\rm had}$ has been observed at LEP.
This quantity is particularly well suited for detecting
non--SM vertex corrections
since the leading QCD corrections cancel, to leading order, in the ratio.
However, since a shift in the couplings of the $b$  will also affect
observables such as $R_Z = \Gamma_{\rm had}/\Gamma_{\ell^+\ell^-}$
and $\sigma^0_{\rm had}$,
it is important to analyze all the precision electroweak data
in a systematic fashion for possible signatures of such
corrections.

\section{Sensitivity to oblique and non--oblique corrections}

In the standard renormalization scheme where $\alpha$, $G_\mu$, and $m_Z$
are used as input to fix the theory, electroweak observables get their
dependence on oblique corrections through the $\rho$ parameter and
$\seff$.   If we denote the contribution of new physics to these
two quantities as $\drho$ and $\ds$, respectively, we have
\beqa
\rho  & = & [\rho]_{\rm SM} + \drho,   \cr
\seff & = & [\seff]_{\rm SM} + \ds,
\eeqa
where $[\cal{O}]_{\rm SM}$ denotes the Standard Model prediction of
the observable $\cal{O}$.

The left and right handed couplings of the $b$ quark to the $Z$
are given by
%\beqa
%\glb  =  [\glb]_{\rm SM} + \frac{1}{3}\,\ds + \dglb, \cr
%\grb  =  [\grb]_{\rm SM} + \frac{1}{3}\,\ds + \dgrb,
%\eeqa
\beq
\glb  =  [\glb]_{\rm SM} + \frac{1}{3}\,\ds + \dglb, ~~~
\grb  =  [\grb]_{\rm SM} + \frac{1}{3}\,\ds + \dgrb,
\eeq
where we have included possible non--oblique corrections from
new physics, $\dglb$ and $\dgrb$.
Assuming that there are no other non--oblique corrections from
new physics, we can calculate the dependence of various
observables on $\drho$, $\ds$, $\dglb$, and $\dgrb$.
It is convenient to define the following linear combinations
of $\dglb$ and $\dgrb$:
\beqa
\xib   & \equiv & (\cos\phi_b)\dglb - (\sin\phi_b)\dgrb,  \cr
\zetab & \equiv & (\sin\phi_b)\dglb + (\cos\phi_b)\dgrb,
\label{xizetadef}
\eeqa
where $\phi_b \equiv \tan^{-1}|\grb/\glb| \approx 0.181$.
By expanding $\Gamb$ and
$A_b \equiv [(\glb)^2-(\grb)^2]/[(\glb)^2+(\grb)^2]$
about the point $\ds=\xib=\zetab=0$,
we find
\beqa
\Gamb
& = & [ \Gamb ]_{\rm SM}
      \left(  1 + \drho - 1.25\,\ds  - 4.65\,\xib
      \right),  \cr
A_b
& = & [ A_b ]_{\rm SM}
      \left( 1 - 0.68\,\ds - 1.76\,\zetab
      \right).
\label{Abzeta}
\eeqa
All the other observables get their dependence on
$\dglb$ and $\dgrb$ through either $\Gamb$ or
$A_b$ so they will depend on either $\xib$ or $\zetab$, but not both.
The observables that depend on $\Gamb$ are:
\beqa
\Gamma_Z & = & [\Gamma_Z]_{\rm SM}
               \left( 1 + \drho - 1.06\,\ds - 0.71\,\xib
               \right),   \cr
\sig0 & = & [\sig0]_{\rm SM}
               \left( 1 + 0.11\,\ds + 0.41\,\xib
               \right),   \cr
R_Z \;\equiv\; \Gamma_{\rm had}/\Gamma_{\ell^+\ell^-}
    & = & [R_Z]_{\rm SM}
               \left( 1 - 0.85\,\ds - 1.02\,\xib
               \right),   \cr
R_b \;\equiv\; \Gamma_{b\bar{b}}/\Gamma_{\rm had}
    & = & [R_b]_{\rm SM}
               \left( 1 + 0.18\,\ds - 3.63\,\xib
               \right),   \cr
R_c \;\equiv\; \Gamma_{c\bar{c}}/\Gamma_{\rm had}
    & = & [R_c]_{\rm SM}
               \left( 1 - 0.35\,\ds + 1.02\,\xib
               \right).
\label{Rxidep}
\eeqa
Note that only $\Gamma_Z$  depends on $\drho$. All of the other
observables can be expressed as ratios of widths, so that
the $\rho$ dependence cancels
between numerator and denominator.
We will ignore $\Gamma_Z$
in the following in order to keep the number of parameters
at a manageable level.
In an analogous way, we find
\beq
A_{\rm FB}^b = \frac{3}{4} A_e A_b
             = [A_{\rm FB}^b]_{\rm SM}
               \left( 1 - 55.7\,\ds - 1.76\,\zetab
               \right).
\label{AFBbzeta}
\eeq

The relationship between our parameters and others
that have appeared in the literature is as follows.
The parameter $\epsilon_b$ introduced in Ref.~1 is
related to $\dglb$ by
\beq
\epsilon_b = [\epsilon_b]_{\rm SM} - 2\dglb.
\eeq
The parameters $\delta_{bV}$ and $\eta_b$ introduced in Ref.~2
are related to $\xib$ and $\zetab$ by
\beqa
\delta_{bV} & = & [\delta_{bV}]_{\rm SM} - 4.65\,\xib,  \cr
\eta_b      & = & [\eta_b]_{\rm SM}      - 1.76\,\zetab.
\eeqa
%

%\end{document}
%\bye

\section{Determination of $\xib$ and $\zetab$}

In order to constrain $\xib$ and $\zetab$, we must first compute nominal
Standard Model values for the various observables.  This in turn requires
that we specify nominal values for the top and Higgs masses.
In the following, we use $m_t = 175$~GeV and $m_H = 300$~GeV.
It is also necessary to specify the value of $\alpha_s$
used in computing the QCD corrections. Here we will use
$\alpha_s = 0.120 \pm 0.006$,
which is the value determined from
hadronic event shapes, jet rates, and energy correlations \cite{ALPHAS}.
We use this value rather than the $0.123 \pm 0.006$ determined using
lineshape data because it is independent of the $Z$ lineshape
parameters we will be using in this analysis.
For the top and Higgs masses given above, the Standard Model
predictions for the relevant observables are summarized in Table 1,
together with the most recent experimental determinations \cite{GLASGOW}.
The errors on $[\sig0]_{\rm SM}$ and $[R_Z]_{\rm SM}$ are due to
the uncertainty in $\alpha_s$.
\begin{table}
\begin{center}
\caption{Experimental measurements and Standard Model predictions for
	various observables}
\begin{tabular}{ccc}
\hline\hline
Observable & Experiment & SM prediction \\
\hline
$\seff$    & $0.2317\pm 0.0007$ (LEP) $0.2294\pm 0.0010$ (SLD)  & 0.2320 \\
$\sig0$    & $41.49\pm 0.12$ (nb)      & $41.43 \pm 0.03 $ \\
$R_Z$      & $20.795\pm 0.040$         & $20.74 \pm 0.04 $ \\
$R_b$      & $0.2202\pm 0.0020$        & $0.2157$          \\
$R_c$      & $0.1583\pm 0.0098$        & $0.1711$          \\
$A^b_{\rm FB}$ & $0.0967\pm 0.0038$     & $0.0957$          \\
$A_b$      &   $0.99\pm 0.14$          & $0.934$           \\
\hline
\end{tabular}
\end{center}
\end{table}

%
%\beqa
%[\seff]_{\rm SM}        & = & 0.2319             \cr
%[\sig0]_{\rm SM}        & = & 41.44   \pm  0.03  \quad({\rm nb})  \cr
%[R_Z]_{\rm SM}          & = & 20.72   \pm  0.04  \cr
%[R_b]_{\rm SM}          & = & 0.2152             \cr
%[R_c]_{\rm SM}          & = & 0.1713             \cr
%[A_{\rm FB}^b]_{\rm SM} & = & 0.0958    	 \cr
%[A_b]_{\rm SM}          & = & 0.934
%\label{SMpred}
%\eeqa
%
%
%\beqa
%\seff         & = &  0.2317 \pm 0.0007 \quad({\rm LEP}) \cr
%\seff         & = &  0.2294 \pm 0.0010 \quad({\rm SLD}) \cr
%\sig0         & = & 41.49   \pm 0.12   \quad({\rm nb})  \cr
%R_Z           & = & 20.795  \pm 0.040  \cr
%R_b           & = &  0.2202 \pm 0.0020 \cr
%R_c           & = &  0.1583 \pm 0.0098 \cr
%A_{\rm FB}^b  & = &  0.0967 \pm 0.0038 \cr
%A_b           & = &  0.99\phantom{00} \pm 0.14\phantom{00} \quad({\rm SLD})
%\label{EXP}
%\eeqa
%
The LEP value of $\seff$ is the average over the {\it leptonic}
asymmetries only;  since the $b\bar b$ asymmetries are sensitive to vertex
corrections as well as shifts in the value of $\seff$,  they should be
handled separately.

\begin{center}
\psfig{file=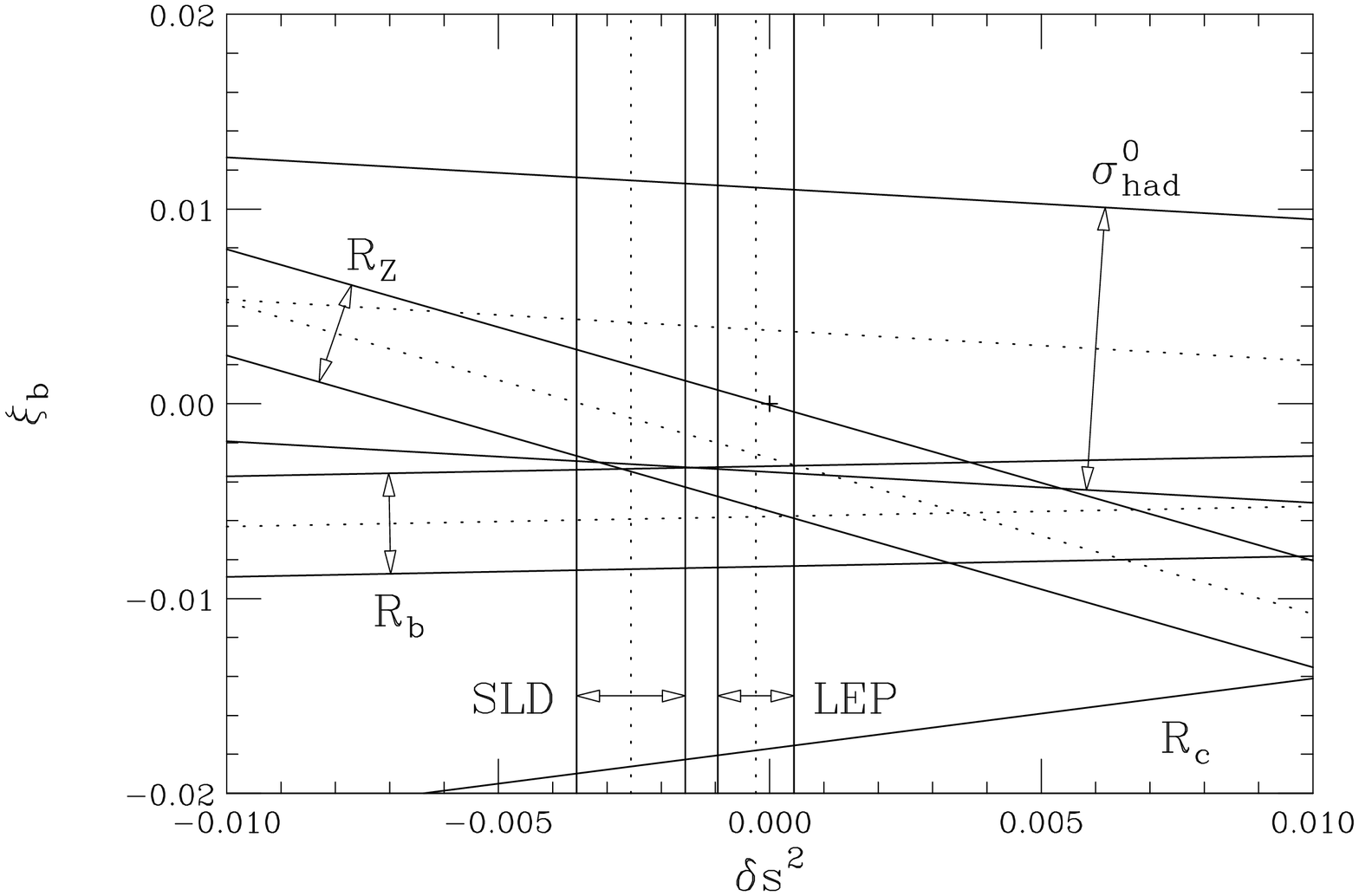,angle=90,width=4in}
{\small {\bf Fig. 1.}  The 1--$\sigma$ limits placed on $\xib$
and $\ds$.}
\end{center}

\vspace{0.5cm}

\begin{center}
\psfig{file=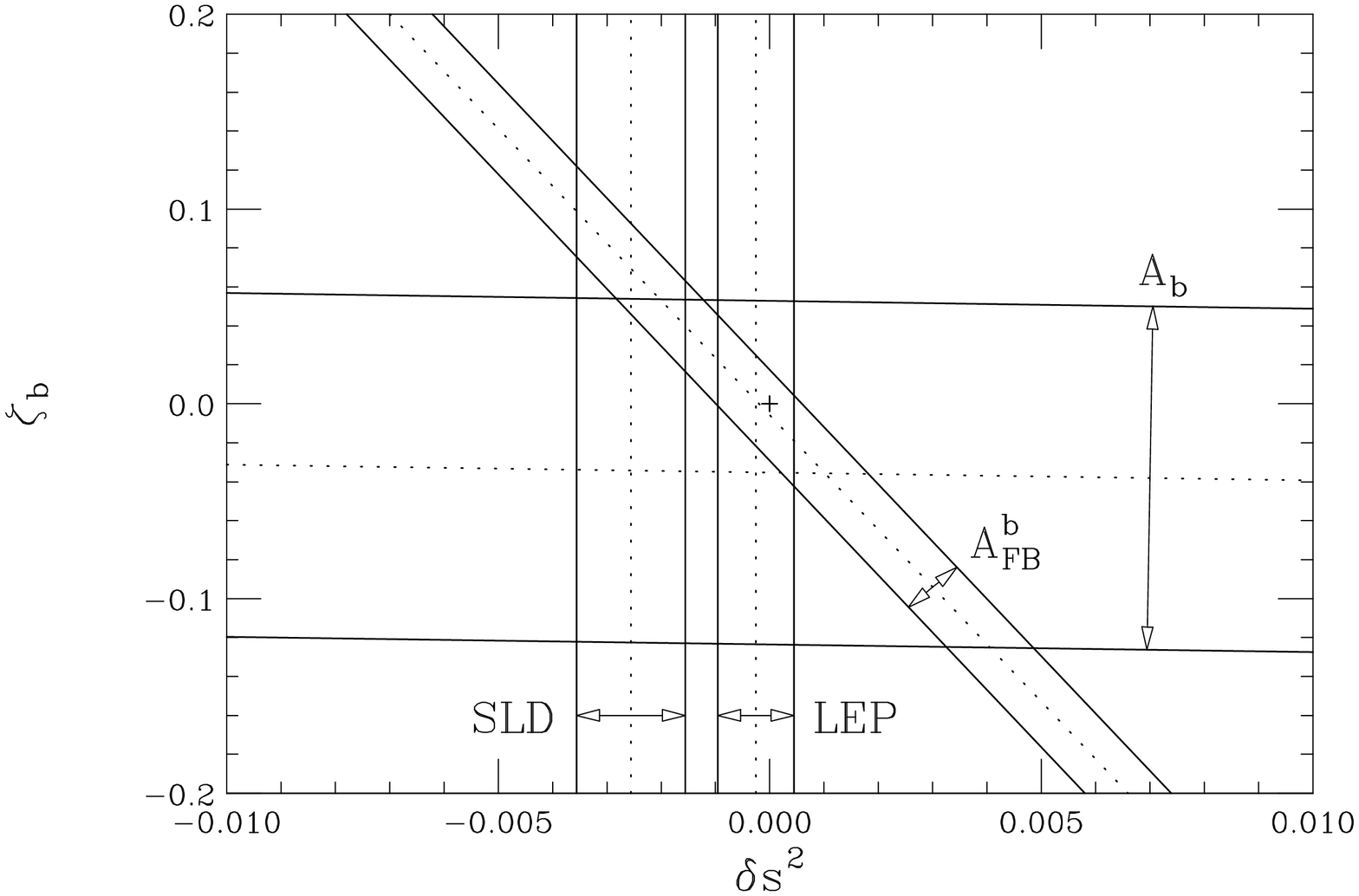,angle=90,width=4in}
{\small {\bf Fig. 2.}  The 1--$\sigma$ limits placed on $\zetab$ and $\ds$.}
\end{center}

The constraints imposed by the various observables are illustrated in
Figures 1 and 2.  In Fig.~1, we show the experimentally preferred
$1-\sigma$ bands
in the $\ds-\xib$ plane, and in Fig.~2 we show the corresponding figure
for the $\ds-\zetab$ plane.
%The bands illustrate the $1-\sigma$ uncertainties
%on the various constraints.

A fit to the data with $\ds$, $\xib$, and $\zetab$ as parameters,
including the correlation of -0.4 between $R_b$ and $R_c$, yields
\beqa
\ds    & = & -0.0009 \pm 0.0006,  \cr
\xib   & = & -0.003\phantom{0}  \pm 0.002,   \cr
\zetab & = & \phantom{-}0.018\phantom{0} \pm 0.027.
\label{ThreeDfit}
\eeqa
The 2--dimensional projections of the allowed regions onto the
$\ds$--$\xib$ and $\ds$--$\zetab$ planes are
shown in Figs.~3 and 4.

\begin{center}
\psfig{file=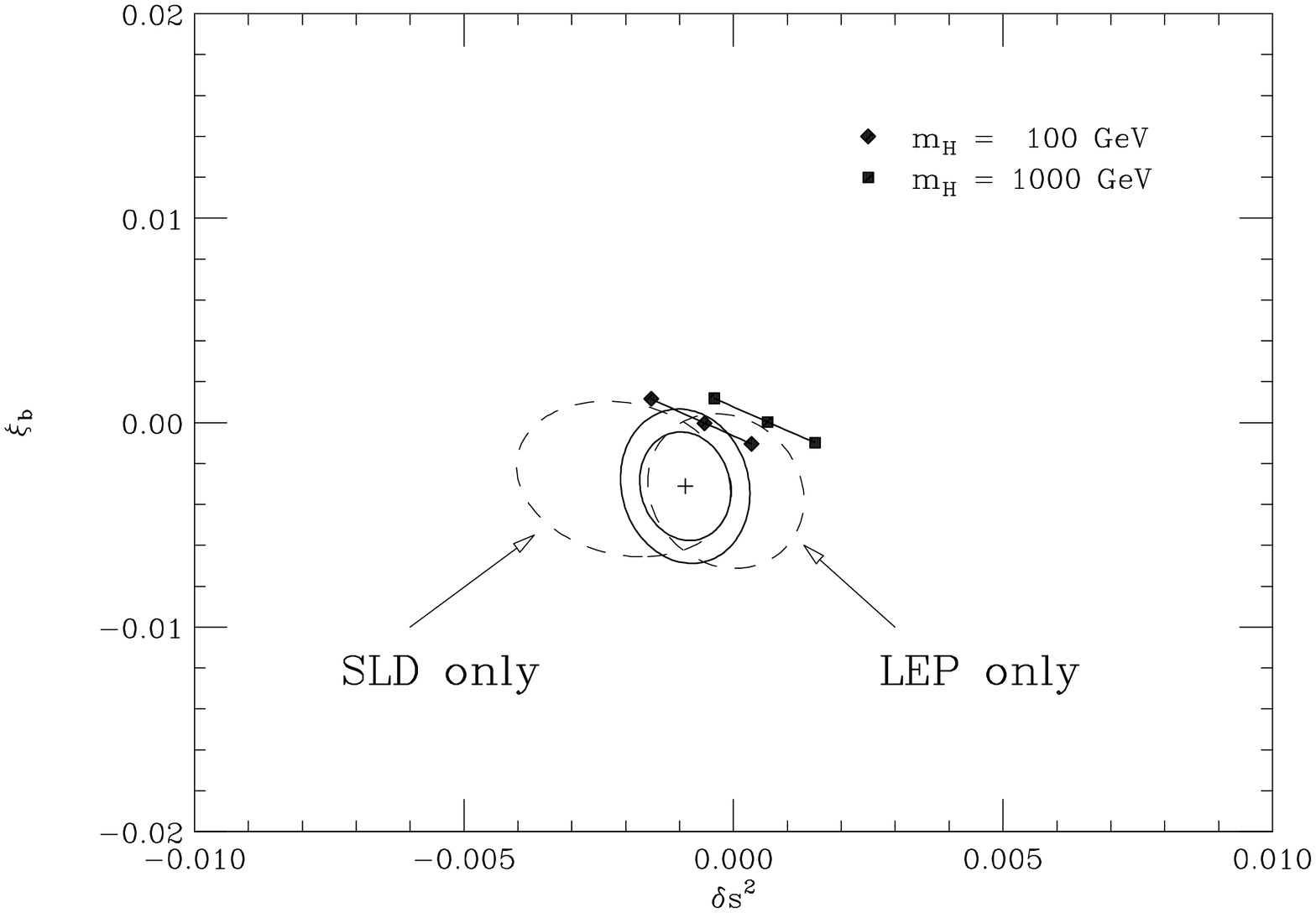,angle=90,width=4in}
{\small {\bf Fig. 3.}
The 68\% and 90\% confidence limits on $\xib$ and $\ds$.
Dashed contours show the positions of the 90\% limit when only the
LEP or SLD value of $\seff$ is used.  The SM points are plotted for
$m_t = 150$, 175, and 200 GeV.  Larger $m_t$ correspond to smaller
$\ds$.}
\end{center}

\vspace{0.5cm}

\begin{center}
\psfig{file=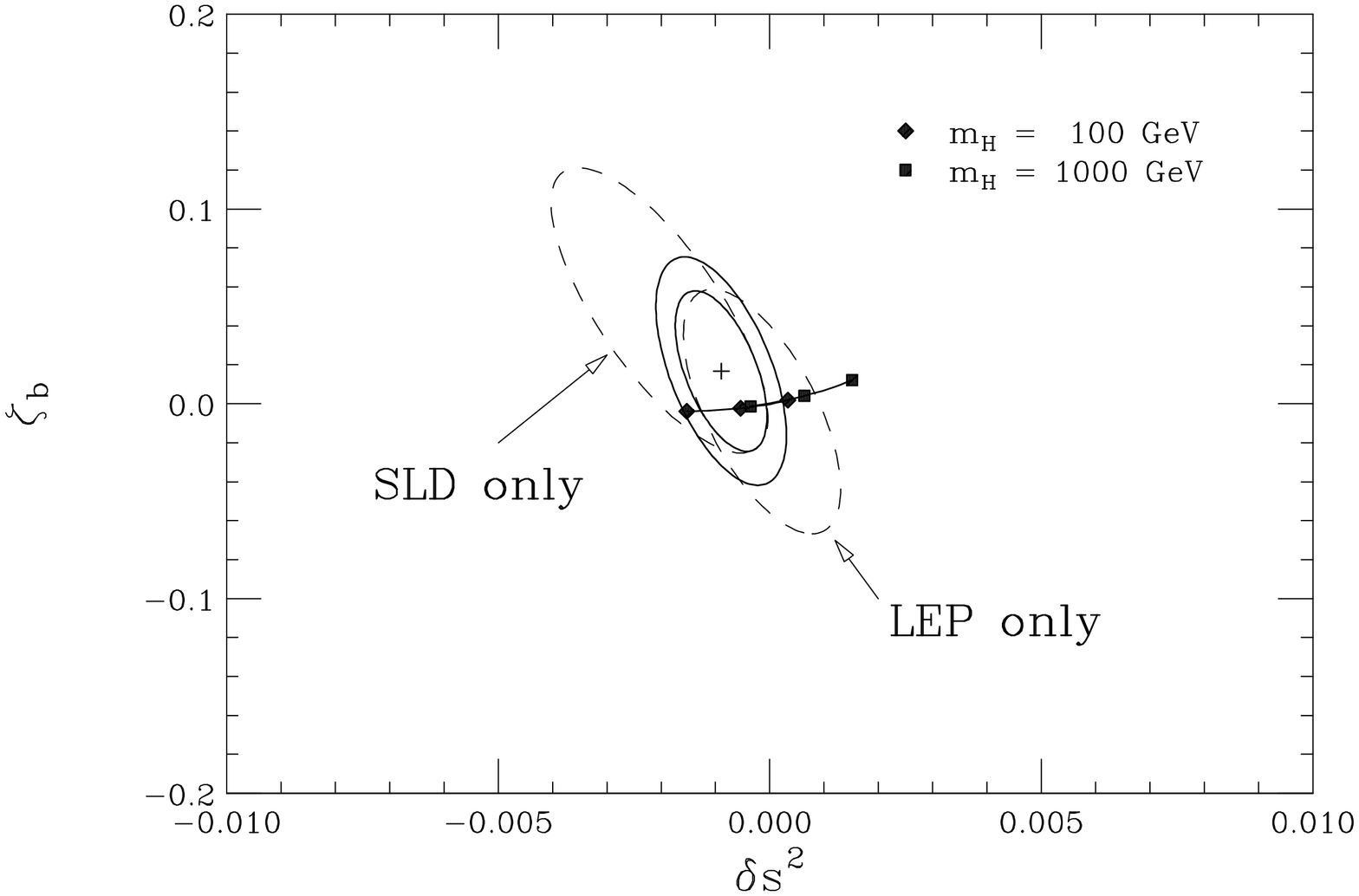,angle=90,width=4in}
{\small {\bf Fig. 4.}
The 68\% and 90\% confidence limits on $\zetab$ and $\ds$.
The meaning of the dashed contours and SM points are the same as in Fig. 3.}
\end{center}

In terms of $\dglb$ and $\dgrb$, Eq.~\ref{ThreeDfit} translates into
\beq
\dglb  = -0.000 \pm 0.005,~~~ \dgrb  =\phantom{-} 0.018 \pm 0.027.
\label{Results3}
\eeq
We see from this that the left--handed coupling of the $b$ is very tightly
constrained by present data, while the right--handed coupling is more
weakly constrained.  This leaves considerable freedom for models
containing extra right--handed gauge bosons or extended Higgs sectors, which
would tend to modify the right--handed coupling of the $b$.  It is also
important to note that many observables, in addition to $R_b$, are
sensitive to shifts in the couplings of the $b$.

\vspace{0.5cm}
{\bf \noindent Acknowledgements \hfil}
\vspace{0.4cm}

We would like to thank M. Peskin, T. Rizzo, J. Hewett,
and T. Onogi for helpful discussions.
This work was supported by
the United States Department of Energy under
Contract Number DE--AC02--76CH03000, and in part
under Grant Number DE--FG02--90ER40560.

\bibliographystyle{unsrt}

\end{document}